\documentclass[12pt, preprint]{aastex}

\newcommand{\bi}[1]{{\mathbf{#1}}}
\newcommand{\bs}[1]{{\mathbf{#1}}}

\def\D{\bi{D}}
\def\beps{\bs{w}}

\def\bS{\bi{S}}

\def\A{\bi{A}}

\def\Sg{\bs{\Sigma}}
\def\Sgi{\Sg^{-1}}
\def\t{^{\dag}}
\def\like{\mathcal{L}}

\def\F{\mathbf{F}}

\def\WMAP{\emph{WMAP}}

\begin{document}

\newcommand{\hinshaw}{hinshaw:2006}
\newcommand{\page}{page:2006}
\newcommand{\spergel}{spergel:2006}
\newcommand{\jarosik}{jarosik:2006}

\title{Point source power in three-year \emph{Wilkinson Microwave
    Anisotropy Probe} data}

\author{
K. M. Huffenberger\altaffilmark{1,2,3}, 
H.\ K.\ Eriksen\altaffilmark{2,3,4,5}, 
F. K. Hansen\altaffilmark{4,5}
}

\altaffiltext{1}{email: huffenbe@jpl.nasa.gov}
\altaffiltext{2}{Jet Propulsion Laboratory, 4800 Oak  Grove Drive, Pasadena CA 91109} 
\altaffiltext{3}{California Institute of Technology, Pasadena, CA  91125} 
\altaffiltext{4}{Institute of Theoretical Astrophysics, University of Oslo, P.O.\ Box 1029 Blindern, N-0315 Oslo, Norway}
\altaffiltext{5}{Centre of Mathematics for Applications, University of Oslo, P.O.\ Box 1053 Blindern, N-0316 Oslo}

\date{Received - / Accepted -}

\begin{abstract}
Using a set of multifrequency cross-spectra computed from the three
year \WMAP\ sky maps, we fit for the unresolved point source
contribution.  For a white noise power spectrum, we find a Q-band
amplitude of $A = 0.011 \pm 0.001$ $\mu$K$^2$ sr (antenna
temperature), significantly smaller than the value of $0.017 \pm
0.002$ $\mu$K$^2$ sr used to correct the spectra in the \WMAP\
release.  Modifying the point source correction in this way largely
resolves the discrepancy \citet{eriksen:2006} found between the \WMAP\
V- and W-band power spectra.  Correcting the co-added \WMAP\ spectrum
for both the low-$\ell$ power excess due to a sub-optimal
likelihood approximation---also reported by \citet{eriksen:2006}---and 
the high-$\ell$ power deficit due to over-subtracted point 
sources---presented in this letter---we find that the net effect in terms of
cosmological parameters is a $\sim0.7\sigma$ shift in $n_{\textrm{s}}$
to larger values: For the combination of \emph{WMAP}, BOOMERanG and
Acbar data, we find $n_{\textrm{s}} = 0.969 \pm 0.016$, lowering the
significance of $n_{\textrm{s}} \ne 1$ from $\sim2.7\sigma$ to $\sim
2.0\sigma$.
\end{abstract}



\section{Introduction}

The results of \emph{Wilkinson Microwave Anisotropy Probe} have made
an inestimable impact on the science of cosmology, highlighted by the
very recent release of the three year data: maps, power spectra, and
consequent cosmological analysis
\citep{\jarosik,\page,\hinshaw,\spergel}.  Precisely because these
results play so prominent a role, it is important to check and recheck
their consistency.


Recently \cite{eriksen:2006} reanalyzed the \WMAP\ three year
temperature sky maps, and noted two discrepancies in the \WMAP\ power
spectrum analysis.  On large angular scales there is a small power
excess in the \emph{WMAP} spectrum (5--10\% at $\ell \lesssim 50$),
primarily due to a problem with the 
 likelihood
approximation used by the \emph{WMAP} team. On small angular scales,
an unexplained systematic difference between the V- and W-band spectra
(few percent at $\ell \gtrsim 300$) was found.  In this Letter, we
suggest this second discrepancy is at least partially due to an
excessive point source correction in the \WMAP\ power spectrum.


\section{Data}
\label{sec:data}

The \emph{WMAP} temperature data \citep{\hinshaw} are provided as ten sky maps 
observed at five frequencies between 23 and 94 GHz, pixelized using the
HEALPix\footnote{http://healpix.jpl.nasa.gov} scheme with 3 million ($\sim$ 7\arcmin-size) 
pixels per map.
Here we consider the
Q-band (41 GHz), V-band (61 GHz), and W-band (94 GHz) channels since
these have the least galactic foreground contamination, but only
V- and W-bands for the
cosmological parameter analysis.

We account for the (assumed circularly symmetric) beam profile
of each channel independently, adopting the Kp2 sky cut as our
mask. This excludes 15.3\% of the sky including all resolved point
sources.  To deal with contamination outside the mask, we simply use
the foreground template corrected maps provided on the LAMBDA
website\footnote{http://lambda.gsfc.nasa.gov/}.
The noise is modeled as uncorrelated,
non-uniform, and Gaussian with an RMS given by $\sigma_{0,i} /
\sqrt{N_{\textrm{\scriptsize obs},i}(p)}$. Here $\sigma_{0,i}$ is the noise per
observation for channel $i$, and $N_{\textrm{\scriptsize obs},i}(p)$ is the number
of observations in pixel $p$.

\section{Methods}
\label{sec:methods}

\subsection{Power spectrum estimation}

We estimate power spectra with the pseudo-$C_{\ell}$ MASTER method
\citep{hivon:2002}, which decouples the mode correlations in a
noise-corrected raw quadratic estimate of the power spectrum computed
on the partial sky.
%
Following \citet{hinshaw:2003}, we include only cross-correlations
between channels in our power spectrum estimates.

Considering each of three years, three bands, and the number of
differencing assemblies per band (two for Q-/V- and four for W-band),
276 individual cross-spectra are available for analysis.  Each of
these is computed to ${\ell_{\rm max} = 1024}$.  The V- and W-band spectra
have been verified against spectra provided by the WMAP team, but the 
Q-band spectra (computed the same way) were not available for comparison.
For the point source amplitude analysis, we bin the power spectra into
ten bins ($\ell=2$--$101$, $102$--$201$,\dots, $902$--$1001$) in order
to increase the signal-to-noise ratio and decrease the number of bins
(and thus the computation time).  The corresponding error bars are
computed using a Fisher approximation and similarly binned. 

\subsection{Point source amplitude estimation}

For our main result, we marginalize over the CMB power and estimate a
single amplitude for the point source spectrum by the method we
discuss below.  We also compute the amplitude in $\ell$-bins, but for
brevity omit the details, which are similar.
We model the ensemble averaged cross-spectra as the sum of the two components,
$\langle  C^\bi{i}_l \rangle = C^{\bi{i},\rm CMB}_l +  C^{\bi{i},\rm src}_l$,
showing explicitly the contribution from each part of the signal.
Here the multipole bin is denoted by $l$ and the cross-correlation
pair by $\bi{i} = {(i_1)(i_2)} = $ (W1yr1)(W2yr3), (Q1yr2)(V1yr2),
{\it etc.}  No auto-power spectra are included, so noise subtraction
is unnecessary.
We marginalize over the CMB spectrum, which we denote by
$C_l^{\rm{CMB}}$.  The window functions for each differencing assembly
pair are $\beps = \{ w^\bi{i}_{ll'} \}$, which we later consider in
terms of a matrix.  The contribution to a cross-spectrum from the CMB
signal is thus
$C^{\bi{i},\rm CMB}_l =  \sum_{l'} w^\bi{i}_{ll'} C^{\rm{CMB}}_{l'}$ (in thermodynamic temperature units).
The spectra in this application are already beam-deconvolved, so the window functions
$w^\bi{i}_{ll'} = \delta_{ll'}$ are trivial.
We denote the amplitude of the unresolved point source power spectrum
by $A$. This amplitude relates to the cross-spectra via the frequency
and shape dependence vector $\bS = \{ S^\bi{i}_{l} \}$,
\begin{eqnarray}
\nonumber
C_l^{\bi{i},\rm src} &=& A S^\bi{i}_l  \\ 
 S^\bi{i}_{l} &=&  w^\bi{i}_{ll} \ \left( \frac{\nu_{i_1}}{ \nu_0} \right)^{\beta}\ 
 K\left( x(\nu_{i_1})\right) 
 \left( \frac{\nu_{i_2} }{ \nu_0} \right)^{\beta} \ 
 K\left( x(\nu_{i_2}) \right) \nonumber \\
K(x) &=& \frac{\left(\exp(x) - 1 \right)^2}{x^2 \exp(x)}.
\end{eqnarray}
Here the cross-spectrum $\bi{i}$ has channels at $\nu_{i_1}$ and
$\nu_{i_2}$, and $x(\nu) = {h \nu}/{ k_B T_{\rm CMB}}$.  The units of
$A$ are antenna temperature squared times solid angle and the function
$K(x)$ converts from antenna temperature to thermodynamic
temperature.  Thus, we assume that the radio sources are spatially
uncorrelated (and therefore have a white noise spectrum) and have a
power-law frequency dependence using antenna temperature units.
Note that well-resolved point sources have already been masked from
the maps before the evaluation of the cross-spectra, and $A$ therefore
represents unresolved sources only. However, we may only directly measure
the frequency dependence for the resolved sources.  For these,
\citet{bennett:2003} found $\beta=-2.0$, and  following
\citet{hinshaw:2003,\hinshaw} we take the same even for the unresolved
sources. We choose $\nu_0=40.7$ GHz (Q-band) as our reference
frequency.

We organize the binned cross-spectra $C^\bi{i}_l$ into a data vector
$\D = \{ C^\bi{i}_l \}$.  We use a Gaussian model for the likelihood
$\like$ of the power spectrum, appropriate at high $\ell$:
\begin{equation}
\like \propto \exp \left\{ -\frac{1}{2}
\left[ \D -  \langle \D \rangle\right]\t \Sgi \left[ \D -
 \langle \D \rangle \right]) \right\},
\end{equation}
where the covariance $\Sg = \langle (\D-\langle \D \rangle)(\D-\langle
\D \rangle)\t \rangle$ can be written as $\Sg = \{
\Sigma^{\bi{ii'}}_{ll'} \}$.  Here we assume the covariance is
diagonal both in multipole and cross-spectrum.  An appendix of
\cite{huffenberger:2004} derives an unbiased estimator for this type
of problem, generalizing the point source treatment of
\citet{hinshaw:2003}.  Here the estimators are equivalent, and  
result in a linear estimate for $A$,
denoted $\bar A$, and its covariance $\Sigma^A$:
\begin{eqnarray} \label{eqn:est}\nonumber 
\bar A &\equiv& (\bS\t\F\bS)^{-1} \bS\t \F \D \\ 
\Sigma^A &\equiv& (\bS\t \F \bS)^{-1},
\end{eqnarray}
where we have defined the auxiliary matrix
\begin{eqnarray} \label{eqn:estaux}\nonumber
\F &\equiv& \Sgi -  \Sgi \beps\left(\beps\t \Sgi \beps\right)^{-1} \beps\t \Sgi.
\end{eqnarray}
In this notation, we consider $\D$ and $\bS$ as column vectors with a
single index $\bi{i}l$, and $\beps$ as a matrix with indices $\bi{i}l$
and $l'$.  Matrices $\Sg$ and $\F$ have indices $\bi{i}l$ and
$\bi{i}'l'$. This estimator marginalizes out the CMB, a conservative treatment which assumes nothing but the 
frequency dependence.  
%
To compute the amplitude in bins, we redefine $A$ as $\A$, a vector of
the amplitudes, with $C^{\bi{i},\rm src}_l = ( \A \cdot \bS
)^{\bi{i}}_l$, modifying $\bS$ for each component to lend power only
to appropriate multipole bins.

\section{Results}
\label{sec:results}

\subsection{Point source spectrum amplitude}

\begin{figure}
\plotone{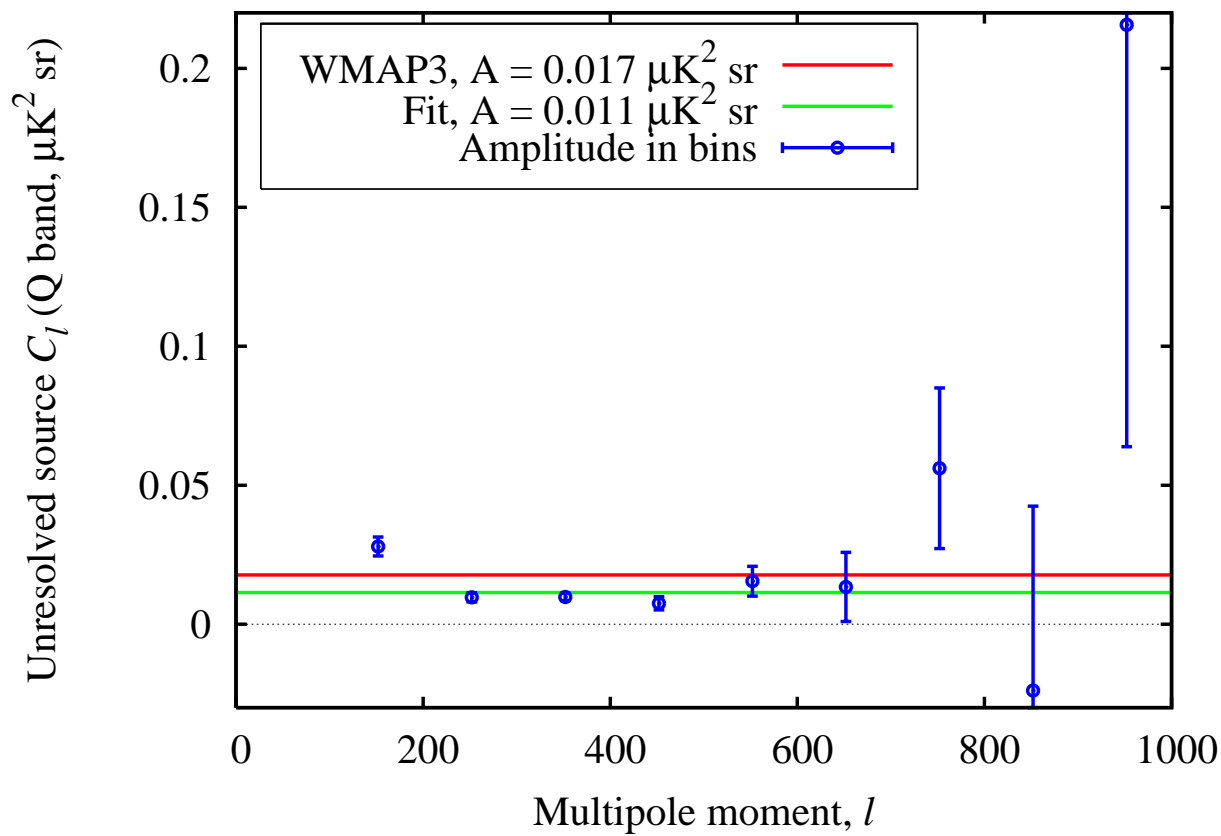}
\caption{The point source power spectrum, fit from \WMAP\ Q-, V-, and
W- bands.  The lowest $\ell$ bin is not plotted, because the error
bars span the entire range of the plot, and it has little statistical
influence.  The 1-parameter fit for a flat spectral shape is also
shown, as well as the point source amplitude from \cite{\hinshaw}.}
\label{fig:source_powspec}
\end{figure}

Using the method described in the previous section, we find a point
source amplitude of $A = 0.011 \pm 0.001$ $\mu$K$^2$ sr, significantly
less than the \WMAP\ value of $A = 0.017 \pm 0.002$ $\mu$K$^2$ sr
\citep{\hinshaw}.  Computing the spectrum in bins (Figure
\ref{fig:source_powspec}), we see that the source power spectrum is
best measured at $100<\ell<600$.   To evaluate goodness-of-fit, 
we compute
$\chi^2 = \sum_{l\textrm{\scriptsize\ bin}}(A_l -
A)^2/\sigma^2_l  = 36.6$  for 9 degrees of freedom.  All of the
discrepancy in our fit arises from a single high bin at
$\ell=102$--$201$, which has $\Delta \chi^2 = 27.2$.  
This bin is so different that we suspect that it is not detecting point source
power alone, but perhaps some residual foreground.  
We leave a rigorous investigation of this anomalous bin to later work,
 leaving it in our analysis here.  If we were to exclude it, the other nine
bins are consistent with a flat power spectrum at $A = 0.011$
$\mu$K$^2$ sr---with $\chi^2 = 9.4$ for 8 remaining degrees of
freedom---though they would prefer a somewhat smaller value for
$A$.  For the \WMAP\ amplitude, we measure $\chi^2 = 86.5$ for 9 degrees of freedom.
This large discrepancy is puzzling because our method should be equivalent to the \WMAP\ method.


\subsection{Angular CMB power spectrum}

The net effect of the lower unresolved point source amplitude on the
co-added \WMAP\ CMB power spectrum may be computed in terms of a
weighted average of corrections for individual cross-spectra
(V$\times$V, V$\times$W, and W$\times$W, respectively). Following the
construction of \WMAP's spectrum, for $\ell < 500$ the correction is
given by a uniform average over the 137 individual cross-spectrum
corrections; for $\ell \ge 500$ it is given as an inverse noise
weighted average \citep{\hinshaw}. In this Letter we approximate the
latter with the inverse variance of the power spectrum coefficients
computed from 2500 simulations for each cross-spectrum individually,
but do not account for correlations between different cross-spectra.

\begin{figure}
\plotone{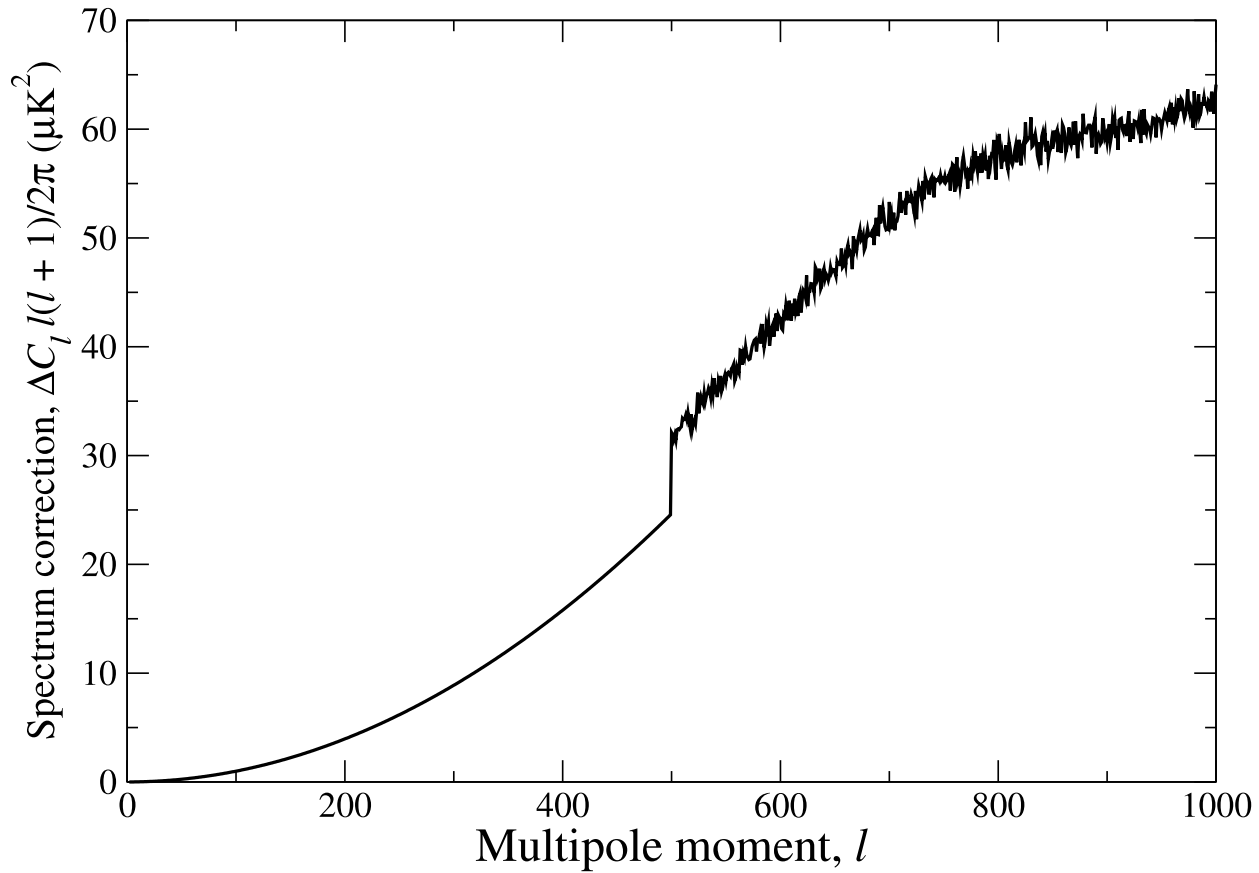}
\caption{The net difference [$(C_{\ell}^{\rm new} -
C_{\ell}^{\rm old})\ell(\ell+1)/2\pi$] in the final co-added \WMAP\ spectrum due
to the new and smaller point source amplitude. The sharp break at $\ell=500$ is
due to different weighting schemes, and the smaller fluctuations at
high $\ell$'s are due to a finite number of Monte Carlo simulations
for noise estimation. } \label{fig:spec_corr}
\end{figure}

\begin{figure}
\epsscale{0.7}
\plotone{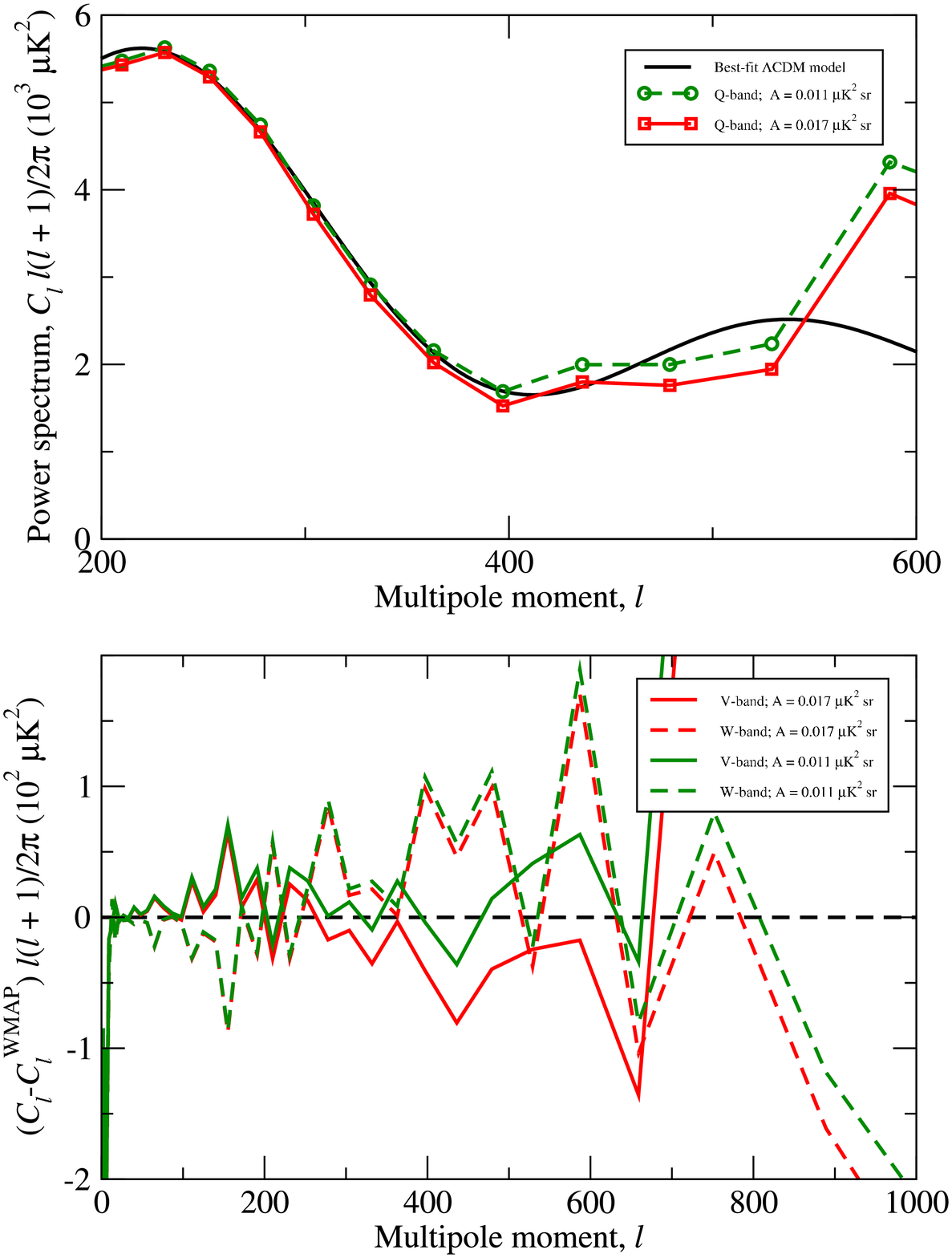}
\caption{The impact on the Q-, V-, and W-band power spectra of the revised
  point source correction.  The top panel shows the Q-band power
  spectrum with the \WMAP\ point source correction (red) and the
  correction in this work (green dashed), plotted with the \WMAP\ best-fit
  $\Lambda$CDM spectrum.  Particularly at $\ell < 400$ where noise is lower, this highlights the point source over-subtraction using the \WMAP\ correction.  The bottom panel shows the V-band (solid) and W-band
  (dashed) power spectra minus the co-added \WMAP\ temperature
  spectrum \citep{\hinshaw}, computed with the \WMAP\ point source
  correction (red) and the correction in this work (green).  The V-
  and W- bands are internally more consistent with the revised source
  correction.}\label{fig:powspec}
\end{figure}

The net power spectrum correction is shown in Figure
\ref{fig:spec_corr}.  We show the Q-band spectra, corrected by each
point source amplitude, in the top panel of Figure \ref{fig:powspec},
and compare the V- and W-band spectra in the bottom panel.

One of two issues pointed out by \citet{eriksen:2006} was a
discrepancy between the V- and W-bands at $\ell \gtrsim 250$
significant at about $3\sigma$. This is seen by comparing the two red
curves in the bottom panel of Figure \ref{fig:powspec}. However,
applying the lower point source correction raises the V-band spectrum
by 10--$50\,\mu\textrm{K}^2$ in this range but the W-band by only a
few $\mu\textrm{K}^2$.  Effectively, about $20\,\mu\textrm{K}^2$ of
the previous $65\,\mu\textrm{K}^2$ average difference is thus removed,
reducing the significance of the difference from 3 to $2\sigma$,
compared to 2500 simulations. A small difference is still
present, and may warrant further investigation, but is no longer
striking.  This gives us confidence that our point source correction
is the more consistent than the \WMAP\ value.

\subsection{Cosmological parameters}

To assess the impact of this new high-$\ell$ correction on
cosmological parameters, we repeat the analysis described by
\citet{eriksen:2006} using the CosmoMC package \citep[][which also
gives the parameter definitions]{lewis:2002} and a modified version of
the \WMAP\ likelihood code \citep{\hinshaw}. First, at $\ell\le30$ the
\WMAP\ likelihood is replaced with a Blackwell-Rao Gibbs
sampling-based estimator
\citep{jewell:2004,wandelt:2004,eriksen:2004,chu:2005}, and second,
the bias correction shown in Figure \ref{fig:spec_corr} is added to
the co-added \WMAP\ spectrum. The results from these computations are
summarized in Table \ref{tab:parameters}.

\begin{figure}
\plotone{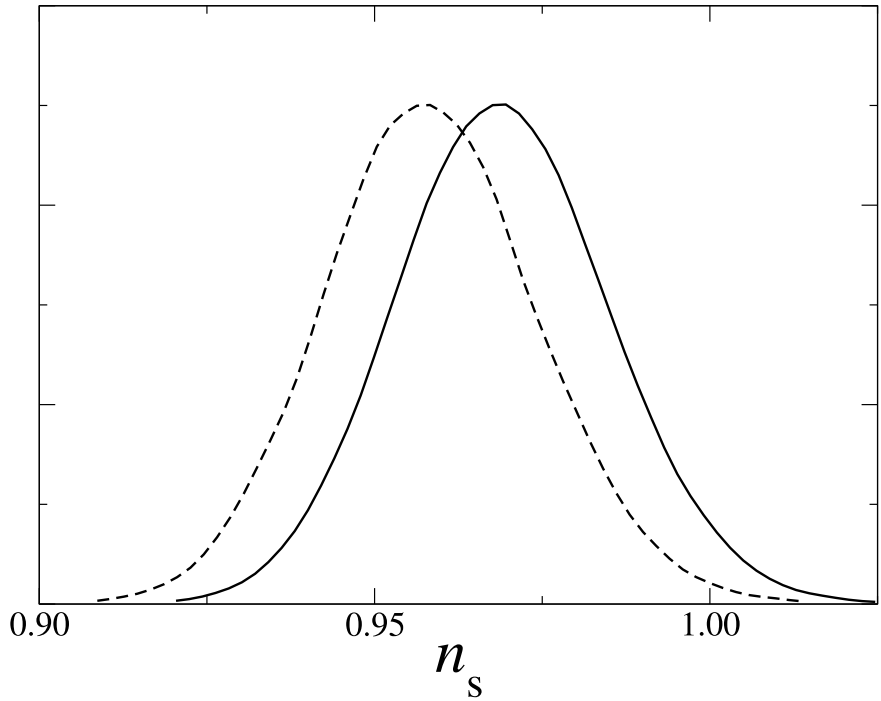}
\caption{Marginalized posterior distributions for the spectral index
  $n_{\textrm{s}}$ computed with the combination of \emph{WMAP},
  BOOMERanG and Acbar data, both with the WMAP likelihood code as
  provided (dashed) and after applying a low-$\ell$ estimator
  correction and a high-$\ell$ point source correction (solid).}
  \label{fig:n_s}
\end{figure}

As reported by \citet{eriksen:2006}, the most notable effect of the
low-$\ell$ estimator bias in the \WMAP\ data release was a
$\sim0.4\sigma$ shift in $n_{\textrm{s}}$ to lower values, increasing
the nominal significance of $n_{\textrm{s}} \ne 1$. In Table
\ref{tab:parameters} we see that the over-estimated point source
amplitude causes a similar effect by lowering the high-$\ell$ spectrum
too much.  Correcting for both of these effects, the spectral index is
$n_{\textrm{s}} = 0.969\pm0.016$ for the combination of \emph{WMAP},
BOOMERanG \citep{montroy:2005,piacentini:2005,jones:2005} and Acbar
\citep{kuo:2004} data, or different from unity by only
$\sim2\sigma$. The marginalized distributions both with and without
these corrections are shown in Figure \ref{fig:n_s}. The
other cosmological parameters change little.  For reference, the
best-fit (as opposed to marginalized) parameters for this case are
$\{\Omega_{\textrm{b}}h^2, \Omega_{\textrm{c}}h^2, h, \tau,
n_{\textrm{s}}, \log (10^{10} A_{\textrm{s}})\} = (0.0225, 0.108,
0.732, 0.919, 0.967, 3.05)$.

\section{Conclusions}
\label{sec:conclusions}

Using a combination of cross-spectra of maps from the Q-, V-, and
W-bands of \WMAP\ three year data, we fit for the amplitude of the
power spectrum of unresolved point sources in Q-band, finding $A =
0.011 \pm 0.001$ $\mu$K$^2$ sr.  This fit has significantly less power
than the fit used to correct the \WMAP\ final co-added power spectrum
used for cosmological analysis.

We compute and apply the proper point source correction, noting the
corrected V- and W-bands are more consistent than before.  The
improper point source correction conspires with a low-$\ell$ estimator
 bias to impart a spurious tilt to the \WMAP\ temperature
power spectrum.  With the revised corrections, we find evidence for
spectral index $n_\textrm{s} \neq 1$ at only $\sim 2\sigma$, while
other parameters remain largely unchanged.

\begin{acknowledgements}
  We thank Gary Hinshaw for useful discussions and comments. We
  acknowledge use of HEALPix software \citep{gorski:2005} for deriving some 
results in this paper. We
  acknowledge use of the Legacy Archive for Microwave Background Data
  Analysis (LAMBDA). This work was partially performed at the Jet
  Propulsion Laboratory, California Institute of Technology, under a
  contract with the National Aeronautics and Space Administration. HKE
  acknowledges financial support from the Research Council of Norway.
\end{acknowledgements}


\clearpage

\begin{deluxetable}{cccc}
\tablewidth{0pt} 
\tabletypesize{\small} 
\tablecaption{Cosmological parameters\label{tab:parameters}}
\tablecolumns{3}
\tablehead{Parameter & \emph{WMAP} & Low-$\ell$ and PS corrected
}

\startdata

\cutinhead{\emph{WMAP} data only}
 $\Omega_{\textrm{b}}\,h^2$    & $0.0222 \pm 0.0007$ & $0.0223\pm0.0007$ \\ 
 $\Omega_{\textrm{m}}$    & $0.241 \pm 0.036$ & $0.244\pm 0.035$ \\ 
 $\log(10^{10}A_{\textrm{s}})$    & $3.019 \pm 0.067$ &
 $3.039\pm 0.068$ \\ 
 $h$    & $0.731 \pm 0.033$ & $0.730\pm 0.032$ \\ 
 $n_{\textrm{s}}$    & $0.954 \pm 0.016$ & $0.966\pm 0.016$ \\ 
 $\tau$    & $0.090 \pm 0.030$ & $0.090\pm 0.030$ \\

\cutinhead{\emph{WMAP} + Acbar + BOOMERanG}
 $\Omega_{\textrm{b}}\,h^2$    & $0.0225 \pm 0.0007$ & $0.0225\pm0.0007$ \\ 
 $\Omega_{\textrm{m}}$    & $0.239 \pm 0.031$ & $0.240\pm 0.031$\\ 
 $\log(10^{10}A_{\textrm{s}})$    & $3.030 \pm 0.064$ &
 $3.045\pm 0.065$ \\ 
 $h$    & $0.737 \pm 0.029$ & $0.738\pm 0.030$\\ 
 $n_{\textrm{s}}$    & $0.958 \pm 0.016$ & $0.969\pm 0.016$\\ 
 $\tau$    & $0.091 \pm 0.030$ & $0.091\pm 0.030$
\enddata

\tablecomments{Comparison of marginalized parameter results obtained
  from the \emph{WMAP} likelihood (second column) and from the
  \emph{WMAP} + Blackwell-Rao hybrid, applying the low-$\ell$
  estimator and high-$\ell$ point source corrections (third column).}

\end{deluxetable}

\clearpage

\clearpage

\clearpage

\clearpage

\end{document}